\documentstyle[12pt]{article}

\begin{document}

\def\text#1{\mbox{#1}}
\def\textit#1{{\it#1}}
\def\textbf#1{{\bf#1}}
\def\mathit#1{{\it#1}}
\def\mathbf#1{{\bf#1}}
\def\mathcal#1{{\cal#1}}
\def\stackunder#1#2{#2_{#1}}

\setlength{\textheight}{1.2\textheight}
\begin{titlepage}
\nopagebreak
\begin{center}
{\large \bf Acceleration of quantum fields}\\
\end{center}
\vfill
\begin{center}
{\bf Marc-Thierry Jaekel$^a$ and Serge Reynaud$^b$} \\
\end{center}

\begin{flushleft}
$(a)$Laboratoire de Physique Th\'{e}orique
 de l'ENS\footnote{Unit\'e
propre du Centre National de la Recherche Scientifique, \\
associ\'ee \`a l'Ecole Normale Sup\'erieure et \`a l'Universit\'e
de Paris Sud.}(CNRS-UPS),\\
24 rue Lhomond, F75231 Paris Cedex 05\\
$(b)$Laboratoire Kastler Brossel\footnote{Unit\'e
de l'Ecole Normale Sup\'erieure et de l'Universit\'e Pierre et Marie Curie,\\
 associ\'ee au Centre National de la Recherche
Scientifique.}(UPMC-ENS-CNRS), case 74,\\
4 place Jussieu, F75252 Paris Cedex 05\\
\end{flushleft}
\vfill

\begin{abstract}
We analyze the transformation of quantum fields under conformal coordinate
transformations from inertial to accelerated frames, in the simple case of
scalar massless fields in a two-dimensional spacetime, through the
transformation of particle number and its spectral density. Particle number
is found to be invariant under conformal coordinate transformations to
uniformly accelerated frames, which extends the property already known for
vacuum. Transformation of spectral density of particle number exhibits a
redistribution of particles in the frequency spectrum. This redistribution
is determined by derivatives of phase operators with respect to frequency,
that is by time and position operators defined in such a manner that the
redistribution of particles appears as a Doppler shift which depends on
position in spacetime, in conformity with Einstein equivalence principle.
\end{abstract}
\begin{flushleft}
\vfill
{\normalsize LPTENS 95/30\\}
\qquad {\it to appear in} Brazilian Journal of Physics,

\qquad  special issue on Quantum Optics, ed. L. Davidovich.
\end{flushleft}

\end{titlepage}

\section{Introduction}

Lorentz invariance of electromagnetism lies at the heart of the theory of
relativity\cite{Einstein1905}. This is true not only for the classical
theory of electromagnetism, but also for the quantum theory. In particular,
invariance of vacuum fluctuations under Lorentz transformations is needed to
ensure that mechanical effects of these fluctuations preserve the relativity
of uniform motion in empty space.

In contrast with these universally accepted ideas, the interplay between
quantum fields and accelerated frames has been the object of much debate.
Since Einstein\cite{Einstein1907}, accelerated frames are commonly
represented by using Rindler changes of coordinates\cite{Rindler} between
inertial and accelerated frames. These transformations do not preserve the
propagation equations of electromagnetic fields. Light rays appear curved in
accelerated frames while frequencies undergo a shift during light propagation%
\cite{Einstein1907}. Such a representation of accelerated frames also
results in a transformation of vacuum into a thermal bath\cite{Unruh}. This
idea has apparently been easily accepted because of its association with the
most spectacular effect predicted by quantum field theory in curved
spacetime, namely thermal particle creation due to curvature\cite{Hawking}.
It is nevertheless clear that accelerated frames and curved spacetime are
completely different physical problems, from the point of view of general
relativity. Furthermore, the notion of particle number plays a central role
in the interpretation of quantum field theory, and the fact that it is not
preserved in accelerated frames leads to weighty paradoxes for quantum theory%
\cite{UnruhWald}. These difficulties also raise doubts about the
significance of the Einstein equivalence principle in the quantum domain.
Such a principle indeed relies on the very notion of particle number, as
well as on the interpretation of frequency change from inertial to
accelerated frames as a Doppler shift depending on position in spacetime\cite
{Einstein1907}. If vacuum or 1-photon states may not be defined in a
consistent manner in inertial and accelerated frames, it might appear
hopeless to attribute any significance to such a principle in the quantum
domain.

In the present paper, we show that the interplay between acceleration and
quantum fields may be analyzed in a consistent manner which allows to extend
the Einstein equivalence principle to the quantum domain. Our approach makes
use of the conformal symmetry of quantum theory of massless fields, like a
scalar field in two-dimensional (2D) spacetime or the electromagnetic field
in four-dimensional (4D) spacetime. For the sake of simplicity, we will
restrict our attention here to the 2D case.

It has been known for a long time that the symmetry related to inertial
motions, associated with Lorentz transformations, can be extended for
massless fields to a larger group which includes conformal transformations
to accelerated frames\cite{BC}. Light rays remain straight lines with such a
representation of accelerated frames while frequencies are now preserved
during light propagation. These transformations are known to fit\cite{Hill}
the relativistic definition of uniformly accelerated motion\cite{Born}. It
has also been shown that vacuum remains unchanged under conformal
transformations to accelerated frames\cite{QSO95}. Here, we extend the
latter property by demonstrating the invariance of total particle number
under such transformations. This proves the consistency of a point of view
which maintains invariance of vacuum and particle number for inertial and
accelerated observers.

When a spectral decomposition of particle number is introduced and the
transformation of spectral density from inertial to accelerated frames
analyzed, field phase operators make an appearance. We will show that the
resulting expressions correspond to quantum definitions for position in
spacetime\cite{NW} which comply with the requirement of the Einstein
equivalence principle for the interpretation of acceleration on quantum
fields in terms of Doppler shifts. It is well known that the definition of
phase operators, which may be considered as conjugated to the number
operators\cite{Dirac27}, leads to ambiguities\cite{Louisell,PhaseOp}. A lot
of work has been devoted to cure these ambiguities (see\cite{BP86} and
references therein). However, the conclusions that we will reach in the
present paper will essentially be unaffected by these difficulties.

\section{Conformal coordinate transformations}

In a two-dimensional (2D) spacetime, a free massless scalar field $\phi
\left( t,x\right) $ is the sum of two counterpropagating components:
\begin{equation}
\phi \left( t,x\right) =\varphi \left( u\right) +\psi \left( v\right) \qquad
u=t-x\qquad v=t+x
\end{equation}
We use natural spacetime units ($c=1$); $t$ is the time coordinate, $x$ is
the space coordinate, $u$ and $v$ are the two light-cone variables.

In the 2D case, conformal coordinate transformations are those
transformations which act separately on the two light-cone variables, and
they are specified by arbitrary functions $f$ and $g$ describing the
relations between such variables in the two reference systems:
\begin{equation}
\overline{u}=f\left( u\right) \qquad \qquad \overline{v}=g\left( v\right)
\label{transfuv}
\end{equation}
The field transformation under conformal coordinate transformations is
defined through:
\begin{equation}
\varphi \rightarrow \overline{\varphi }\qquad \qquad \varphi \left( u\right)
=\overline{\varphi }\left( \overline{u}\right)  \label{transfphi}
\end{equation}
{}From now on, we consider only one ($\varphi $) of the two counterpropagating
components; the other one ($\psi $) can be dealt with in exactly the same
way.

As well-known in Quantum Field Theory, the field transformation under
coordinate transformations may be considered as generated by linear forms of
the stress tensor, that is also quadratic forms of the fields\cite{Itzykson}%
. In order to give explicit forms of these generators in the spectral
domain, we introduce the Fourier components $\varphi \left[ \omega \right] $
of the field $\varphi \left( u\right) $ according to the general definition:
\begin{equation}
\varphi \left( u\right) =\int \frac{d\omega }{2\pi }\varphi \left[ \omega
\right] e^{-i\omega u}
\end{equation}
These components are related to the standard annihilation and creation
operators:
\begin{eqnarray}
\varphi \left[ \omega \right] &=&\sqrt{\frac \hbar {2\left| \omega \right| }}%
\left( \theta \left( \omega \right) a_\omega +\theta \left( -\omega \right)
a_{-\omega }^{\dagger }\right) \\
\left[ a_\omega ,a_{\omega ^{\prime }}\right] &=&\left[ a_\omega ^{\dagger
},a_{\omega ^{\prime }}^{\dagger }\right] =0  \label{commutAA} \\
\left[ a_\omega ,a_{\omega ^{\prime }}^{\dagger }\right] &=&2\pi \delta
\left( \omega -\omega ^{\prime }\right)  \label{commutAAz}
\end{eqnarray}
where $\theta $ is the Heaviside function and $\delta $ the Dirac
distribution. The commutation relations of the Fourier components of the
field are given by:
\begin{equation}
\left[ \varphi \left[ \omega \right] ,\varphi \left[ \omega ^{\prime
}\right] \right] =\frac{\pi \hbar }\omega \delta \left( \omega +\omega
^{\prime }\right)  \label{commutphi}
\end{equation}

We now define the generating function $T\left[ \omega \right] $:
\begin{equation}
T\left[ \omega \right] =\int \frac{d\omega ^{\prime }}{2\pi }\omega ^{\prime
}\left( \omega +\omega ^{\prime }\right) \varphi \left[ -\omega ^{\prime
}\right] \varphi \left[ \omega +\omega ^{\prime }\right]  \label{Tomega}
\end{equation}
as the Fourier transform of the stress tensor\footnote{%
In this definition appear symmetric products of the field operators, rather
than normally ordered products. As a consequence, the generating function
describes the stress tensor associated with vacuum, as well as the stress
tensor associated with particles.}:
\begin{equation}
T\left( u\right) =\left( \partial _u\varphi \left( u\right) \right) ^2
\end{equation}
We then introduce the generators $T_k$ as the coefficients of the Taylor
expansion of the generating function ($k$ a positive integer):
\begin{equation}
T_k=\left\{ \left( -i\partial _\omega \right) ^kT\left[ \omega \right]
\right\} _{\omega =0}=\int u^kT\left( u\right) du  \label{Tk}
\end{equation}
The commutators of these quantities with the field are obtained as:
\begin{eqnarray}
\left[ T\left[ \omega \right] ,\varphi \left[ \omega ^{\prime }\right]
\right] &=&-\hbar \left( \omega +\omega ^{\prime }\right) \varphi \left[
\omega +\omega ^{\prime }\right] \\
\left[ T_k,\varphi \left[ \omega \right] \right] &=&-\hbar \left( -i\partial
_\omega \right) ^k\left\{ \omega \varphi \left[ \omega \right] \right\}
\end{eqnarray}
The latter relation precisely fits the action upon the field of an
infinitesimal conformal coordinate transformation. Denoting:
\begin{equation}
\delta \varphi \left[ \omega \right] =\frac \varepsilon {i\hbar }\left[
T_k,\varphi \left[ \omega \right] \right]
\end{equation}
with $\varepsilon $ an infinitesimal real number, one indeed deduces:
\begin{equation}
\delta \varphi \left( u\right) =-\varepsilon u^k\partial _u\varphi \left(
u\right)
\end{equation}
This corresponds to equations (\ref{transfuv}-\ref{transfphi}) with an
infinitesimal coordinate transformation:
\begin{eqnarray}
\overline{u} &=&u+\delta f\left( u\right) \\
\delta \varphi \left( u\right) &\equiv &\overline{\varphi }-\varphi =-\delta
f\left( u\right) \partial _u\varphi \left( u\right) \\
\delta f\left( u\right) &=&\varepsilon u^k
\end{eqnarray}
Notice that the generating function $T\left[ \omega \right] $ may also be
associated with an infinitesimal coordinate transformation:
\begin{equation}
\delta f\left( u\right) =\varepsilon \exp \left( i\omega u\right)
\end{equation}
This does not correspond to a real coordinate tranformation which would
necessarily involve opposite values of the frequency\footnote{%
We may emphasize that we are dealing with coordinate transformations of real
Minkowski spacetime, here represented by real light-cone variables. The
generating function (\ref{Tomega}), that is the Fourier transform of the
field stress tensor, is not hermitian but satisfies $T\left[ \omega \right]
^{\dagger }=T\left[ -\omega \right] $. The generators $T_k$ are defined in
equation (\ref{Tk}) as hermitian operators, in contrast with common
definitions of generators in Conformal Field Theory in 2D spacetime\cite{CFT}%
.}.

In order to recover the known commutation relations for the conformal
generators\cite{CAlg,CFT}, we write the commutator of the generating
function with quadratic forms of the field:
\begin{eqnarray}
\left[ T\left[ \omega \right] ,\varphi \left[ \omega ^{\prime }\right]
\varphi \left[ \omega ^{\prime \prime }\right] \right] &=&-\hbar \{\left(
\omega +\omega ^{\prime }\right) \varphi \left[ \omega +\omega ^{\prime
}\right] \varphi \left[ \omega ^{\prime \prime }\right]  \nonumber \\
&&+\left( \omega +\omega ^{\prime \prime }\right) \varphi \left[ \omega
^{\prime }\right] \varphi \left[ \omega +\omega ^{\prime \prime }\right] \}
\label{QuadraticForms}
\end{eqnarray}
We then deduce the commutator of the generating function evaluated at
different arguments:
\begin{equation}
\left[ T\left[ \omega \right] ,T\left[ \omega ^{\prime }\right] \right]
=\hbar \left( \omega -\omega ^{\prime }\right) T\left( \omega +\omega
^{\prime }\right)  \label{commutTomega}
\end{equation}
A Taylor expansion of this relation provides commutators characteristic of
the conformal algebra (for positive integers $k$)\footnote{%
Note that the original Virasoro algebra is defined with negative and
positive order non-hermitian generators. It is generated by $\frac 1{\hbar
\Omega }T\left[ k\Omega \right] $, where $\Omega $ is a scale frequency. It
also corresponds to a Laurent expansion in variable $u$ of the function $%
T\left( u\right) $ extended to the complex plane\cite{CFT}.}:
\begin{equation}
\left[ T_k,T_{k^{\prime }}\right] =i\hbar \left( k^{\prime }-k\right)
T_{k+k^{\prime }-1}  \label{commutTk}
\end{equation}

\section{Transformation of vacuum}

The conformal coordinate transformations preserve the propagation equation
of massless fields, and therefore their commutators\cite{QSO95}. However,
not all of them preserve vacuum fluctuations.

The vacuum state is defined by specific correlation functions:
\begin{equation}
\left\langle \varphi \left[ \omega \right] \varphi \left[ \omega ^{\prime
}\right] \right\rangle _{\text{vac}}=\theta \left( \omega \right) \theta
\left( -\omega ^{\prime }\right) \left[ \varphi \left[ \omega \right]
,\varphi \left[ \omega ^{\prime }\right] \right]
\end{equation}
$\left\langle ~\right\rangle _{\text{vac}}$ represents a mean value in the
vacuum state. This means that annihilators vanish when applied to the vacuum
state. Using expression (\ref{commutphi}) of the field commutators, one
obtains the correlation function:
\begin{equation}
\left\langle \varphi \left[ \omega \right] \varphi \left[ \omega ^{\prime
}\right] \right\rangle _{\text{vac}}=\theta \left( \omega \right) \frac{\pi
\hbar }\omega \delta \left( \omega +\omega ^{\prime }\right)
\end{equation}
Using transformation (\ref{QuadraticForms}) of field quadratic forms, one
then deduces the transformation of vacuum correlation functions:
\begin{equation}
\left\langle \left[ T\left[ \omega \right] ,\varphi \left[ \omega ^{\prime
}\right] \varphi \left[ \omega ^{\prime \prime }\right] \right]
\right\rangle _{\text{vac}}=\pi \hbar ^2\left( \theta \left( \omega ^{\prime
}\right) -\theta \left( -\omega ^{\prime \prime }\right) \right) \delta
\left( \omega +\omega ^{\prime }+\omega ^{\prime \prime }\right)
\end{equation}
It is also worth writing the transformation of the vacuum stress tensor,
that is of the generating function itself\footnote{%
When written in terms of normally ordered products, commutation relations (%
\ref{commutTomega}) between the generators include a further pure number.
This central charge is determined by equation (\ref{commutTomegaVac}).}:
\begin{equation}
\left\langle \left[ T\left[ \omega \right] ,T\left[ \omega ^{\prime }\right]
\right] \right\rangle _{\text{vac}}=\frac{\hbar ^2\omega ^3}{12}\delta
\left( \omega +\omega ^{\prime }\right)  \label{commutTomegaVac}
\end{equation}
One then demonstrates, through a Taylor expansion of these relations in the
frequency $\omega $, that the vacuum correlation function for field
derivatives $\partial _u\varphi $ (which correspond to Fourier components $%
-i\omega \varphi \left[ \omega \right] $), as well as the vacuum stress
tensor, are preserved by the infinitesimal generators $T_0$, $T_1$ and $T_2$
which respectively describe translations, Lorentz boosts and conformal
transformations from inertial to accelerated frames. This is no longer the
case for the higher-order generators. In particular, the generator $T_3$
changes the vacuum stress tensor in a manner which is consistent with the
dissipative force felt by a mirror moving in vacuum with a non-uniform
acceleration\cite{FDavies,MTJSRqo}.

We thus recover the result of reference\cite{QSO95} for a massless scalar
field theory in a 2D spacetime: the vacuum is not invariant under the large
group of conformal coordinate transformations (equation (\ref{transfuv})
with an arbitrary function $f$). It is invariant only under the smaller
group of transformations generated by $T_0$, $T_1$ and $T_2$. Those
transformations correspond to the particular case of homographic functions%
\footnote{%
Note that the modification of the mean vacuum stress tensor $\left\langle
T\left( u\right) \right\rangle _{\text{vac}}$ under a conformal
transformation associated with the function $f$ is proportional to the
Schwartzian derivative of $f$, which vanishes for homographic transformations%
\cite{CFT,FDavies}.}:
\begin{equation}
\overline{u}=\frac{au+b}{cu+d}\qquad \qquad ad-bc=1
\end{equation}
In the following, we give some results for the large conformal group, but we
focus our attention onto the smaller group of transformations which preserve
vacuum, and particularly onto the action of the acceleration generator $T_2$.

\section{Transformation of particle number operators}

We will denote $n_\omega $ the spectral density of particle number:
\begin{equation}
n_\omega =a_\omega ^{\dagger }a_\omega =\frac{2\omega }\hbar \theta \left(
\omega \right) \varphi \left[ -\omega \right] \varphi \left[ \omega \right]
\end{equation}
The values at different frequencies are commuting quantities:
\begin{equation}
\left[ n_\omega ,n_{\omega ^{\prime }}\right] =0  \label{commutNomega}
\end{equation}
and the commutators with the field may be written from relations (\ref
{commutphi}):
\begin{equation}
\left[ n_\omega ,\varphi \left[ \omega ^{\prime }\right] \right] =2\pi
\left\{ \delta \left( \omega +\omega ^{\prime }\right) -\delta \left( \omega
-\omega ^{\prime }\right) \right\} \varphi \left[ \omega ^{\prime }\right]
\end{equation}

This definition is such that the generator $T_0$, that is the field energy,
has its standard form in terms of number density:
\begin{equation}
T_0=\left\langle T_0\right\rangle _{\text{vac}}+\int_0^\infty \frac{d\omega
}{2\pi }\hbar \omega n_\omega  \label{T0}
\end{equation}
The total number $n$ of particles is defined as the integral of $n_\omega $:
\begin{equation}
n=\int_0^\infty \frac{d\omega }{2\pi }n_\omega  \label{defN}
\end{equation}
The number operators $n_\omega $ are defined for positive frequencies, and
vanish when applied to the vacuum state.

We come now to the main argument of the present paper, that is the
transformation of particle numbers under conformal coordinate
transformations. As an immediate consequence of transformation (\ref
{QuadraticForms}) of field quadratic forms, we deduce the transformation of
the number density:
\begin{eqnarray}
\left[ T\left[ \omega \right] ,n_{\omega ^{\prime }}\right] &=&-2\omega
^{\prime }\theta \left( \omega ^{\prime }\right) \{\left( \omega -\omega
^{\prime }\right) \varphi \left[ \omega -\omega ^{\prime }\right] \varphi
\left[ \omega ^{\prime }\right]  \nonumber \\
&&+\left( \omega +\omega ^{\prime }\right) \varphi \left[ -\omega ^{\prime
}\right] \varphi \left[ \omega +\omega ^{\prime }\right] \}
\label{LomegaNomega}
\end{eqnarray}
We obtain the transformation of the total particle number by an integration:
\begin{equation}
\left[ T\left[ \omega \right] ,n\right] =-\int_0^\omega \frac{d\omega
^{\prime }}\pi \omega ^{\prime }\left( \omega -\omega ^{\prime }\right)
\varphi \left[ \omega -\omega ^{\prime }\right] \varphi \left[ \omega
^{\prime }\right]  \label{LomegaN}
\end{equation}
We then derive the effect of the infinitesimal generators by performing a
Taylor expansion in the frequency $\omega $ of the previous expressions.

The total particle number $n$ is preserved by the generators $T_0$, $T_1$
and $T_2$:
\begin{equation}
\frac 1{i\hbar }\left[ T_0,n\right] =\frac 1{i\hbar }\left[ T_1,n\right]
=\frac 1{i\hbar }\left[ T_2,n\right] =0
\end{equation}
This property is well-known for the translations and Lorentz boosts. The new
result is that a conformal transformation to an accelerated frame also leads
to a redistribution of particles in the frequency domain, without any change
of the total number of particles. It is consistent with the invariance of
vacuum in the homographic group generated by $T_0$, $T_1$ and $T_2$, as
discussed in the previous section. It means that the notion of particle
number is the same for accelerated observers and for inertial ones, provided
that accelerated frames are defined through conformal transformations. For
the other generators $T_{k\geq 3}$, the vacuum is no longer preserved and
the total particle number $n$ is changed\footnote{%
Note that the commutator (\ref{LomegaNomega}) vanishes when applied to the
vacuum state, for arbitrary positive frequencies $\omega $. However, vacuum
and particle numbers are not invariant under generators $T_{k\geq 3}$. These
properties are consistent since, as already mentioned, $T\left[ \omega
\right] $ is not hermitian and real coordinate transformations involve the
generating function $T\left[ \omega \right] $ at negative frequencies as
well as positive ones.}.

We now write the transformation of the spectral density $n_\omega $ of
particle number under the generators $T_0$, $T_1$ and $T_2$ which perserve
the total number $n$. As expected, the number density is unchanged under a
translation:
\begin{equation}
\frac 1{i\hbar }\left[ T_0,n_\omega \right] =0
\end{equation}
but changed under a Lorentz boost:
\begin{equation}
\frac 1{i\hbar }\left[ T_1,n_\omega \right] =\partial _\omega \left\{ \omega
n_\omega \right\}  \label{L1Nomega}
\end{equation}
This latter change is a mere mapping in the frequency domain, associated
with the Doppler shift of the field frequency. We then write the
modification of the spectral density of particle number in a conformal
transformation from an inertial to an accelerated frame:
\begin{eqnarray}
\frac 1{i\hbar }\left[ T_2,n_\omega \right] &=&2\partial _\omega \left\{
\omega m_\omega \right\}  \label{L2Nomega} \\
m_\omega &=&\frac \omega {i\hbar }\theta \left( \omega \right) \left\{
\varphi \left[ -\omega \right] \varphi ^{\prime }\left[ \omega \right]
+\varphi ^{\prime }\left[ -\omega \right] \varphi \left[ \omega \right]
\right\}  \label{Momega}
\end{eqnarray}
The quadratic form $m_\omega $ is hermitian. It may not be rewritten in
terms of the density $n_\omega $ or its derivatives. In other words, the
modification of $n_\omega $ under $T_2$ amounts to a redistribution of
particles in the frequency domain, without any change of the total particle
number, as it was the case for the modification of $n_\omega $ under $T_1$,
but this redistribution is no longer equivalent to a mere mapping of the
density $n_\omega $ in the frequency spectrum. We will show later on that
the expression (\ref{Momega}) may be interpreted as a Doppler shift which
depends on position in spacetime, in conformity with Einstein equivalence
principle.

\section{Quantum phase and phase-time operators}

In the present section, we show how to obtain quantum operators associated
with positions in spacetime.

As a first step in this direction, we introduce operators $e_\omega $ and $%
\delta _\omega $ such that:
\begin{eqnarray}
a_\omega &=&e_\omega \sqrt{n_\omega }  \label{defdelta} \\
a_\omega ^{\dagger } &=&\sqrt{n_\omega }e_\omega ^{\dagger } \\
e_\omega &=&e^{i\delta _\omega }  \label{defdeltaz}
\end{eqnarray}
As well-known, these relations are not sufficient to define phase operators
since annihilators and creators are not modified by a redefinition of the
phases such that:
\begin{equation}
e_\omega \rightarrow e_\omega +\pi _\omega \qquad \qquad \pi _\omega
n_\omega =0
\end{equation}
Various definitions of the phase operators, for example the
Susskind-Glogower definition\cite{PhaseOp} or the Pegg-Barnett definition%
\cite{PB89}, are connected through such redefinitions. We show below that
the properties studied in the present paper may be stated independently of
such ambiguities.

We now list some properties which are satisfied for any operators defined
from relations (\ref{defdelta}-\ref{defdeltaz}); these properties depend
only upon the field commutation relations (\ref{commutAA}-\ref{commutAAz}).
First, the exponential operators $e_\omega $ are commuting variables, like
the number operators (compare with (\ref{commutNomega})):
\begin{equation}
\left[ e_\omega ,e_{\omega ^{\prime }}\right] =0  \label{commutDelta}
\end{equation}
This is also the case for their adjoint operators $e_\omega ^{\dagger }$:
\begin{equation}
\left[ e_\omega ^{\dagger },e_{\omega ^{\prime }}^{\dagger }\right] =0
\label{commutDeltaC}
\end{equation}
The commutation relations between operators $e_\omega $ (or $e_\omega
^{\dagger }$) and the number operators $n_\omega $ satisfy:
\begin{eqnarray}
\left[ n_\omega ,e_{\omega ^{\prime }}\right] \sqrt{n_{\omega ^{\prime }}}
&=&-2\pi e_{\omega ^{\prime }}\delta \left( \omega -\omega ^{\prime }\right)
\sqrt{n_{\omega ^{\prime }}} \\
\sqrt{n_{\omega ^{\prime }}}\left[ n_\omega ,e_{\omega ^{\prime }}^{\dagger
}\right] &=&2\pi \sqrt{n_{\omega ^{\prime }}}e_{\omega ^{\prime }}^{\dagger
}\delta \left( \omega -\omega ^{\prime }\right)
\end{eqnarray}
However, the exponential operators $e_\omega $ do not commute in the general
case with their adjoint operators $e_\omega ^{\dagger }$:
\begin{eqnarray}
e_\omega e_\omega ^{\dagger } &=&1  \label{EEcroix} \\
e_\omega ^{\dagger }e_\omega &=&1-\alpha _\omega \Pi _\omega
\end{eqnarray}
where $\Pi _\omega $ projects onto vacuum for field components at frequency $%
\omega $, and $\alpha _\omega $ is a function of $\omega $ which depends on
the specific definition of the phase operator. It follows that the
exponential operators are not necessarily unitary and, hence, that the phase
operators are not hermitian. One gets for example $\alpha _\omega =1$ in the
Susskind-Glogower definition\cite{PhaseOp}, and $\alpha _\omega =0$ in the
Pegg-Barnett definition, which thus corresponds to hermitian phase operators%
\cite{PB89}. For all definitions, one may nevertheless write
\begin{equation}
\sqrt{n_\omega }e_\omega ^{\dagger }e_\omega =e_\omega ^{\dagger }e_\omega
\sqrt{n_\omega }=\sqrt{n_\omega }
\end{equation}
It follows that simple relations hold for states orthogonal to vacuum, i.e.
states such that the probability for having $n_\omega =0$ vanishes.

We have given definitions of the phase operators for a field having a whole
frequency spectrum, and not only for a monomode field. We are thus able to
deal with frequency variation of the phase operators and, in particular, to
consider the operators $\delta _\omega ^{\prime }$ obtained by
differentiating phases $\delta _\omega $ versus frequency, according to the
Wigner definition of phase-times\cite{Wigner}. A lot of discussions have
been devoted to the significance of such a definition, and of its relation
with time observables which can be measured by various techniques\cite
{TimeMeasurement}. Here, we will emphasize that the operators $\delta
_\omega ^{\prime }$ do not commute with number operators and with energy,
thus providing quantum phase-times.

Since the exponential operators $e_\omega $ commute (see relation (\ref
{commutDelta})), the frequency derivative $\delta _\omega ^{\prime }$ of the
phase may be defined from the frequency derivative $e_\omega ^{\prime }$ of
the exponential operator:
\begin{equation}
e_\omega ^{\prime }=i\delta _\omega ^{\prime }e_\omega =ie_\omega \delta
_\omega ^{\prime }
\end{equation}
It may be defined as well from the adjoint exponential operators $e_\omega
^{\dagger }$:
\begin{equation}
\left( e_\omega ^{\dagger }\right) ^{\prime }=-ie_\omega ^{\dagger }\left(
\delta _\omega ^{\prime }\right) ^{\dagger }=-i\left( \delta _\omega
^{\prime }\right) ^{\dagger }e_\omega ^{\dagger }
\end{equation}
It follows from relation (\ref{EEcroix}) that the phase derivative $\delta
_\omega ^{\prime }$ is an hermitian operator, even for non-hermitian
definitions of the phase $\delta _\omega $:
\begin{equation}
\delta _\omega ^{\prime }=-ie_\omega ^{\prime }e_\omega ^{\dagger
}=ie_\omega \left( e_\omega ^{\dagger }\right) ^{\prime }=\left( \delta
_\omega ^{\prime }\right) ^{\dagger }
\end{equation}

Using these properties and definitions (\ref{defdelta}-\ref{defdeltaz}), we
may now rewrite the definition (\ref{Momega}) of $m_\omega $ as:
\begin{eqnarray}
m_\omega &=&\frac i2\left\{ \left( a_\omega ^{\prime }\right) ^{\dagger
}a_\omega -a_\omega ^{\dagger }a_\omega ^{\prime }\right\}  \label{maaC} \\
m_\omega &=&\sqrt{n_\omega }\delta _\omega ^{\prime }\sqrt{n_\omega }
\label{mphases}
\end{eqnarray}
The quadratic form $m_\omega $ is proportional to the density $n_\omega $,
but also to the operator $\delta _\omega ^{\prime }$ which, as we shall see
in the next section, has properties of a quantum position in spacetime.

The operators $\delta _\omega ^{\prime }$ have been defined from phase
operators, so that they are expected to have non vanishing commutators with
the number operators\cite{Dirac27}. The definition of such commutators is
affected by the ambiguities already discussed\cite{PhaseOp}. We may however
state them in a rigorous manner by evaluating the commutators between the
densities $m_\omega $ and $n_{\omega ^{\prime }}$ (for $\omega >0$ and $%
\omega ^{\prime }>0$):
\begin{equation}
\left[ m_\omega ,n_{\omega ^{\prime }}\right] =-2\pi i\delta ^{\prime
}\left( \omega -\omega ^{\prime }\right) n_\omega  \label{commutmn}
\end{equation}
These relations are unambiguously defined in any quantum state and they are
consistent with Dirac-like commutators in states orthogonal to the vacuum
(states such that $n_\omega \neq 0$):
\begin{equation}
\sqrt{n_\omega }\left[ \delta _\omega ^{\prime },n_{\omega ^{\prime
}}\right] \sqrt{n_\omega }=-2\pi i\delta ^{\prime }\left( \omega -\omega
^{\prime }\right) n_\omega
\end{equation}
To derive this result, we have used relation (\ref{mphases}) and the fact
that $n_\omega $ and $n_{\omega ^{\prime }}$ are commuting variables.

\section{Discussion}

A comparison between the relations (\ref{L1Nomega}) and (\ref{L2Nomega}),
which describe respectively the effect of a Lorentz boost and of a change of
acceleration on the number density, shows that the latter is equivalent to a
Doppler shift of the field frequency which depends on the operator $\delta
_\omega ^{\prime }$. This property appears to be quite close to a quantum
expression of the Einstein equivalence principle, provided that $\delta
_\omega ^{\prime }$ plays the role of a position in spacetime, in
consistency with the Wigner definition of phase-times\cite{Wigner}. The
semiclassical character of the Wigner definition makes its extension to the
definition of a quantum operator difficult. We show now that it is however
possible to write down rigorous quantum statements with $\delta _\omega
^{\prime }$ used like a position in spacetime.

To this aim, we evaluate commutation relations between $\delta _\omega
^{\prime }$ and the energy operator $T_0$. Multiplying equation (\ref
{commutmn}) by frequency $\omega ^{\prime }$ and integrating over $\omega
^{\prime }$, we get (see equation (\ref{T0})):
\begin{equation}
\left[ T_0,m_\omega \right] =i\hbar n_\omega  \label{commutT0m}
\end{equation}
We may also introduce the integral $m$ of the density $m_\omega $, in the
same manner as the total particle number $n$ from the density $n_\omega $:
\begin{equation}
m=\int_0^\infty \frac{d\omega }{2\pi }m_\omega  \label{defM}
\end{equation}
We deduce from the commutator (\ref{commutT0m}):
\begin{equation}
\left[ T_0,m\right] =i\hbar n  \label{commutT0M}
\end{equation}
We notice that the commutation relations between $m$ and the creation and
annihilation operators have a simple form:
\begin{equation}
\left[ m,a_\omega \right] =ia_\omega ^{\prime }\qquad \qquad \left[
m,a_\omega ^{\dagger }\right] =i\left( a_\omega ^{\prime }\right) ^{\dagger }
\label{commutMaaC}
\end{equation}
We now discuss these relations from the point of view of the quantum
definition of positions in spacetime.

We first discuss the spectral relation (\ref{commutT0m}). Since $n_\omega $
is invariant in a translation, we deduce from relation (\ref{mphases}):
\begin{equation}
\sqrt{n_\omega }\left[ T_0,\delta _\omega ^{\prime }\right] \sqrt{n_\omega }%
=i\hbar n_\omega
\end{equation}
For states orthogonal to the vacuum state ($n_\omega \neq 0$), this has the
form of a canonical commutator between $T_0$ and $\delta _\omega ^{\prime }$%
, thus defining $\delta _\omega ^{\prime }$ as a quantum phase-time.

More exactly, $T_0$ is the energy associated with the light-cone variable $u$%
, so that $\delta _\omega ^{\prime }$ has to be interpreted as a quantum
operator $U_\omega $ having this variable $u$ as its classical analog. The
same manipulations applied to the counterpropagating field component $\psi $
would lead to the definition of a quantum variable $V_\omega $ having the
light-cone variable $v$ as its classical analog. Combining these two
variables, it is therefore possible to define time- and space-like
operators:
\begin{equation}
\delta _\omega ^{\prime \ (\varphi )}\equiv U_\omega =\tau _\omega -\xi
_\omega \qquad \qquad \delta _\omega ^{\prime \ (\psi )}\equiv V_\omega
=\tau _\omega +\xi _\omega  \label{lightconeAA}
\end{equation}
which are conjugated to the field energy and momentum:
\begin{equation}
\left[ E,\tau _\omega \right] =i\hbar \qquad \qquad \left[ P,\xi _\omega
\right] =-i\hbar
\end{equation}
defined through:
\begin{equation}
E=T_0^{(\varphi )}+T_0^{(\psi )}\qquad \qquad P=T_0^{(\varphi )}-T_0^{(\psi
)}  \label{lightconeZZ}
\end{equation}
This provides quantum definitions of time and space operators $\tau _\omega $
and $\xi _\omega $, defined at each frequency $\omega $ like the
semiclassical Wigner definitions.

In order to give a more explicit realisation of quantum positions in
spacetime, we now consider the integrated relation (\ref{commutT0M}), in the
particular case of a 1-particle state. As already discussed, the notion of a
number state is preserved in conformal transformations to accelerated
frames; precisely the total particle number $n$ is preserved. In particular,
the definition of a 1-particle state ($n=1$) is the same for accelerated and
inertial observers. For such a state, the commutator (\ref{commutT0M}) now
reads as a canonical commutator between the energy $T_0$ and the operator $m$%
:
\begin{equation}
\left[ T_0,m\right] =i\hbar \qquad \qquad n=1  \label{canocommut}
\end{equation}
This relation may be considered as associating a quantum position to the
1-particle state, precisely one position for each light-cone variable.
Following the same path as from equation (\ref{lightconeAA}) to equation (%
\ref{lightconeZZ}), we may then obtain time and space operators $\tau $ and $%
\xi $ associated with the state.

In fact, the operator $m$ is a generalization for quantum fields of the
Newton-Wigner quantum position\cite{NW}.  This position, initially defined
for a wavefunction, is here extended to 1-particle field states. To make
this point explicit, we represent each 1-particle state by a function $f$ of
frequency or of position:
\begin{equation}
\mid f\rangle =\int_0^\infty \frac{d\omega }{2\pi }f\left[ \omega \right]
\mid \omega \rangle =\int_{-\infty }^\infty du\ f\left( u\right) \mid
u\rangle
\end{equation}
where we have used Dirac-like ket notations for the basis states:
\begin{eqnarray}
\mid \omega \rangle &=& a_\omega ^{\dagger }\mid \text{vac}\rangle  \\
\mid u\rangle &=& \int_0^\infty \frac{d\omega }{2\pi }e^{i\omega u}\mid
\omega \rangle
\end{eqnarray}
Equation (\ref{commutMaaC}) thus means that the operator $m$ may be
represented in the space of functions $f$ either as the differential
operator $\left( -i\partial _\omega \right) $ in the frequency domain, or as
the multiplication by $u$ in the position domain:
\begin{equation}
m\mid f\rangle = -i\int_0^\infty \frac{d\omega }{2\pi }f^{\prime }\left[
\omega \right] \mid \omega \rangle =\int_{-\infty }^\infty du\ u\ f\left(
u\right) \mid u\rangle
\end{equation}
Its spectral density $m_\omega$ can be shown to be related
to its symmetrised product
with particle number density $n_\omega$:
\begin{equation}
{1\over2} \left\{m,n_\omega\right\} \equiv {1\over2} \left( m n_\omega +
n_\omega m \right) = m_\omega + : m n_\omega :
\end{equation}
where $: \quad :$ denotes normal ordering:
\begin{equation}
: m n_\omega : = {i\over2} \int_0^\infty {d\omega^\prime \over 2\pi} \left\{
\left( a_{\omega^\prime}^\prime \right)^\dagger a_\omega^\dagger
a_{\omega^\prime} a_\omega - a_{\omega^\prime}^\dagger
a_\omega^\dagger a_{\omega^\prime}^\prime a_\omega  \right\}
\end{equation}
This normal product vanishes when applied to 1-particle field states, so that
for such states the density $m_\omega$ can be identified, as an operator, with
the
symmetrised product of position $m$ and particle number density $n_\omega$.
Using the commutation relations (\ref{commutMaaC}),
it can then be rewritten under the form (\ref{mphases}) with position $m$
substituted for $\delta _\omega ^{\prime }$:
\begin{equation}
m_\omega = {1\over2} \left\{m,n_\omega\right\} = \sqrt{n_\omega } m
\sqrt{n_\omega } \qquad \qquad n = 1
\end{equation}
Finally, transformations of particle number density
to inertial or accelerated frames take the simple form of
Doppler shifts of the frequency ((\ref{L1Nomega}) and (\ref{L2Nomega})):
\begin{eqnarray}
\frac 1{i\hbar }\left[ T_1,n_\omega \right] &=&\partial _\omega \left\{ \omega
n_\omega \right\}\\
\frac 1{i\hbar }\left[ T_2,n_\omega \right] &=&2\partial _\omega \left\{
\omega \sqrt{n_\omega } \delta _\omega ^{\prime } \sqrt{n_\omega }\right\}
\end{eqnarray}
For 1-particle field states, the last relation can also be written:
\begin{equation}
\frac 1{i\hbar }\left[ T_2,n_\omega \right] =2\partial _\omega \left\{
\omega \sqrt{n_\omega } m \sqrt{n_\omega }\right\} \qquad \qquad n = 1
\end{equation}
This Doppler shift is proportional to the
acceleration and to the Newton-Wigner position of the particle.

We may now summarize the results obtained in this paper. In order to take
advantage of the conformal symmetry of massless field theories, we have
represented accelerated frames by conformal transformations. Invariance of
vacuum under such transformations was already known\cite{QSO95}. We have
demonstrated that total particle number was also invariant, thus proving the
consistency of a point of view where vacuum and number states are the same
for inertial and accelerated observers. In contrast with the common Rindler
representation of accelerated frames discussed in the introduction, this
point of view allows to discuss the effect of acceleration on quantum fields
in terms of a redistribution of particle in the frequency domain. Analyzing
the transformation of spectral density of particle number from inertial to
accelerated frames, we have shown that it may be interpreted in terms of
Doppler shifts depending upon position in spacetime, in conformity with the
Einstein equivalence principle. This position is defined as the frequency
derivative of some phase operators, in analogy with the Wigner definition of
phase-times\cite{Wigner}. In the particular case of 1-particle states, it is
a generalization to Quantum Field Theory of the Newton-Wigner position
operator initially defined for wavefunctions\cite{NW}. Considered as a
whole, these results constitute a step forward in the direction of a
consistent interpretation of the Einstein equivalence principle in the
quantum domain.

\end{document}